# Probability density evolution filter

XU Yazhou[1*]

(1. *School of Civil Engineering, Xi'an University of Architecture and Technology   Shaanxi 710055, China*)

**Abstract**：Based on probability density evolution method (PDEM) and Bayes' law, a new filter strategy is proposed, in which the prior probability of system state of interest is predicted by solving the generalized density evolution equation (GDEE), the posterior probability of system state is then updated in terms of Bayesian formula. Furthermore, a Chebyshev polynomial-based collocation method is employed to obtain numerical solutions of the prior probability. An illustrative example is finally presented to validate the probability density evolution filter (PDEF) in comparison to particle filter (PF) and unscented Kalman filter (UKF). Overall, PDEF exhibits accuracy close to PF without any resampling algorithm.

## 1. Introduction

For many scientific and engineering issues, it is a common concern to predict the current state of a dynamic system on basis of new observation information. In the case of linear system with Gaussian noise, Kalman filter provides the closed optimal solution[1][2]. In consideration of nonlinear system or non-Gaussian noise, extended Kalman filter[3][4], unscented Kalman filter[5][6]and other approximation methods are thereafter proposed[2]. Nevertheless, up to now the most powerful filter method tackling nonlinearity and non-Gaussian is particle filter (PF), it is also named as data assimilation in geophysics. In the framework of particle filter, the joint posterior probability density function (PDF) is approximated by a set of weighted samples, i.e. particles[2]. Firstly, weights of particles at next time are predicted based on the state equations, by Bayesian law the weights are then updated or corrected according to new observation information. Particle filter hence consists of two major steps, prediction and correction. In such a way, the weights are sequentially updated, so do the approximated posterior PDF.

For the original PF it is difficult to sample particles from the desired PDF, a proposal PDF or importance distribution is often introduced to bridge this gap, which results in sequentially importance sampling (SIS) method[7]. Whereas, one often meets so-called weight degeneracy problem, which means when performing SIS the weighted values of few particles are very large while weights of the remaining particles are close to zero. To overcome this drawback, resampling strategy is carried out in each recursive step or after a few steps. Generally, resampling is often implemented to produce a set of particles with equal weight based on the updated posterior PDF, in accordance with sampling importance resampling (SIR) method[8]. Nevertheless, SIR is prone to particle impoverishment, which corresponds to that the resampled particles are duplicated few particles with dominant weights, while other particles with small weights are gradually abandoned. Actually, degeneracy is often associated with impoverishment. In addition, for many high dimensional systems with more than hundred thousands of statement variables, such as dynamic modes in oceanography, meteorology and civil engineering, it is still prohibitive to meet computational demands or complexity to deal with the whole system simultaneously.

In fact, we often focus on a few statement variables or even one key factor, rather than all statement variables. To achieve this aim, in the framework of classical probability theory probability density evolution method (PDEM) [9][10] provides a feasible way. It has been successively used to perform the stochastic response and reliability assessment in many engineering fields [11][12][13][14] [15][16] [17]. The most important merit of PDEM is the capacity to deal with only one variable rather than all system state variables in view of the random event description[18]. In this study, the essential merit of dimension reduction by PDEM [9] is incorporated with Bayes' law, so that we can predict the prior joint PDF of a few dynamic variables by PDEM, and update the prior PDF to obtain corresponding posterior PDF with new observation information based on Bayesian formula. Generally speaking, we can employ this strategy for any arbitrary dimension[20], in practice the case of one variable is often of the most interest.

The rest of this paper is organized as following. The framework of probability density evolution filter (PDEF) is firstly presented in Section 2. Next, a collocation point based spectral method is introduced to solve the generalized density evolution equation in Section 3, and an algorithm of probability density evolution filter in terms of Bayes' formula is summarized to update the prior PDF, so that we can obtain the prior PDF. Section 4 presents an illustrative example to implement PDEF, in comparison to UKF and PF. Finally, some conclusions and suggestions are drawn in Section 5.

## 2. Problem statement of probability density evolution filter (PDEF)

The state equation of a dynamic system could be expressed as

$$\dot{\mathbf{x}} = \mathbf{G}(\mathbf{x}, \mathbf{v}) \tag{1}$$

where $\mathbf{x} = (x_1, x_2, \cdots, x_n)^T \in \mathbb{R}^n$ denotes a $n$-dimensional state vector, $\mathbf{G}: \mathbb{R}^n \times \mathbb{R}^n \to \mathbb{R}^n$ is a linear or nonlinear function modeling the dynamic system. And $\mathbf{v}$ is a $n$-dimensional noise vector, which accounts for model error or other uncertainty. In most cases, there is no a closed-form solution for the state equation with continuous time, we have to resort to numerical methods. Given a proper discretization scheme with equal time step of $\Delta t$, the discrete state equation with additional noise can be written as:

$$\mathbf{x}_k = \mathbf{g}(\mathbf{x}_{k-1}, t_{k-1}) + \mathbf{v}_{k-1} \tag{2}$$

where $k \in \mathbb{N}$ denotes discrete time.

Moreover, it is assumed that the defined dynamic system above can be observed at the $k$th discrete-time step. The $k$th observation is a random vector $\mathbf{y}_k \in \mathbb{R}^m$:

$$\mathbf{y}_k = \mathbf{h}(\mathbf{x}_k, t_k) + \mathbf{w}_k \tag{3}$$

where $\mathbf{h}: \mathbb{R}^n \times \mathbb{R}^m \to \mathbb{R}^m$ is a function connecting the state variables and observation variables of the dynamic system. And $\mathbf{w}_k$ is a $m$-dimensional noise vector, which takes observation error into consideration.

Assume that the problem defined in Eq.(2) is well-posed, the $l$th solution component of

$\mathbf{x}_k$ can be formally written as:

$$x_k^l = g_l(\mathbf{x}_{k-1}, \mathbf{v}_{k-1}), \quad l = 1, \cdots, n \tag{4}$$

Actually, there is no need to obtain the joint probability of all variables simultaneously for many realistic high-dimensional dynamic systems. We often focus on a few state variables much less than all of the entire system. Especially, if one component, e.g. $x_k^l$, is in consideration, we can readily derive the one-dimensional probability density equation as follows.

Denote $\tau \in (t_{k-1}, t_k)$, for simplicity write $x_k^l$ as $x_k$. Thus, $\tau$ is a locally continuous parameter from $t_{k-1}$ to $t_k$. Assume that the posterior PDF $p(x_{k-1}|y_{k-1})$ at $t_{k-1}$ is available. Consider the joint probability domain $\{\Omega_x \times \Omega_{v_{k-1}}\}$, in which $\Omega_x$ denotes the probability space of state variable and $\Omega_{v_{k-1}}$ represents the probability space of noise $\mathbf{v}_{k-1}$. Let $\Omega$ be an arbitrary subdomain and $\partial\Omega$ its boundary. For this preserved stochastic system[10][19], during $d\tau$ the probability inflow through $\partial\Omega$ into $\Omega$ is

$$-\left[p_{xv}(x, \mathbf{v}_{k-1}, \tau|y_{k-1})\dot{x}d\tau\right] \cdot \mathbf{n}dS_x d\mathbf{v}_{k-1} \tag{5}$$

Here, $dS_x$ is the surface element of $\partial\Omega_x$, $\mathbf{n}$ denotes the norm of $\partial\Omega_x$, $\dot{x}$ is the derivative of $x$ with respect to $\tau$, and $d\mathbf{v}_{k-1}$ is the volume element of $\Omega_{v_{k-1}}$.

Meanwhile, the first-order expansion of the joint PDF $p_{xv}(x, \mathbf{v}_{k-1}, \tau+d\tau|y_{k-1})$ at $\tau$ is

$$p_{xv}(x, \mathbf{v}_{k-1}, \tau+d\tau|y_{k-1}) = p_{xv}(x, \mathbf{v}_{k-1}, \tau|y_{k-1}) + \frac{\partial p_{xv}(x, \mathbf{v}_{k-1}, \tau|y_{k-1})}{\partial \tau}d\tau \tag{6}$$

It can be easily seen that the incremental probability during $d\tau$ is $\frac{\partial p_{xv}(x, \mathbf{v}_{k-1}, \tau|y_{k-1})}{\partial \tau}dxd\mathbf{v}_{k-1}d\tau$.

For a preserved stochastic system, there is no additional random factor in $\Omega$ [9]. Therefore, the probability inflow through $\partial\Omega$ into $\Omega$ must be equal to the incremental probability during $d\tau$ [20], it is

$$\int_{\Omega_x \times \Omega_{v_{k-1}}} \frac{\partial p_{xv}(x, \mathbf{v}_{k-1}, \tau|y_{k-1})}{\partial \tau} dxd\mathbf{v}_{k-1}d\tau = -\int_{\partial\Omega_x \times \Omega_{v_{k-1}}} \left[p_{xv}(x, \mathbf{v}_{k-1}, \tau|y_{k-1})\dot{x}d\tau\right] \cdot \mathbf{n}dS_x d\mathbf{v}_{k-1} \tag{7}$$

In terms of Gaussian integration formula, the surface integral on $\partial\Omega_x$ can be transformed into the appropriate volume integral as follows:

$$\int_{\partial\Omega_x \times \Omega_{v_{k-1}}} \left[ p_{xv}(x, \mathbf{v}_{k-1}, \tau | y_{k-1}) \dot{x} d\tau \right] \cdot \mathbf{n} dS_x d\mathbf{v}_{k-1} = \int_{\partial\Omega_x \times \Omega_{v_{k-1}}} \dot{x} \frac{\partial p_{xv}(x, \mathbf{v}_{k-1}, \tau | y_{k-1})}{\partial x} dx d\mathbf{v}_{k-1} d\tau \quad (8)$$

Substitute Eq.(8) into Eq.(7), we have a conserved relation:

$$\int_{\Omega_x \times \Omega_{v_{k-1}}} \left[ \frac{\partial p_{xv}(x, \mathbf{v}_{k-1}, \tau | y_{k-1})}{\partial \tau} + \dot{x} \frac{\partial p_{xv}(x, \mathbf{v}_{k-1}, \tau | y_{k-1})}{\partial x} \right] dx d\mathbf{v}_{k-1} d\tau = 0 \quad (9)$$

Since $\Omega$ is an arbitrary subdomain of $\Omega_x \times \Omega_{v_{k-1}}$, as a consequence the joint PDF $\partial p_{xv}(x, \mathbf{v}_{k-1}, \tau | y_{k-1})$ satisfies with the one-dimensional generalized density evolution equation (GDEE) [9][10]:

$$\frac{\partial p_{xv}(x, \mathbf{v}_{k-1}, \tau | y_{k-1})}{\partial \tau} + \dot{x} \frac{\partial p_{xv}(x, \mathbf{v}_{k-1}, \tau | y_{k-1})}{\partial x} = 0 \quad (10)$$

Correspondingly, the initial condition is:

$$p_{xv}(x, \mathbf{v}_{k-1}, \tau = t_{k-1} | y_{k-1}) = \delta(x) p_x(x_{k-1} | y_{k-1}) p_v(\mathbf{v}_{k-1}) \quad (11)$$

where $p_v(\mathbf{v}_{k-1})$ is the known PDF of noise vector $\mathbf{v}_{k-1}$, $\delta(\cdot)$ is the Dirac's delta function. For more detailed derivation of one-dimensional generalized density evolution equation, we refer readers to [9] [10].

Furthermore, we can obtain the desired prior probability density function $p(x_k | x_{k-1})$ at $t_k$ by integration of the joint PDF $p_{xv}(x, \mathbf{v}_{k-1}, \tau = t_k | y_{k-1})$ over the definition domain of $\mathbf{v}_{k-1}$ as follows,

$$p(x_k | x_{k-1}) = \int_{\Omega_{v_{k-1}}} p_{xv}(x, \mathbf{v}_{k-1}, \tau = t_k | y_{k-1}) dv \quad (12)$$

In this way, we can predict the prior PDF $p(x_k | x_{k-1})$ at $t_k$ through solving GDEE, instead of dealing with the Chapman-Kolmogorov integral equation in PF.

Next, following the well-known Bayesian formulation, we can easily update the prior probability density function $p(x_k | x_{k-1})$ based on the new observation information. That is,

$$p(x_k | y_k) = \frac{q(y_k | x_k) p(x_k | x_{k-1})}{p(y_k | y_{k-1})} \quad (13)$$

where $q(y_k | x_k)$ denotes the likelihood function determined by observation equation in Eq.(3), and $p(y_k | y_{k-1}) = \int_{x_k} p(y_k | x_k) p(x_k | y_{k-1}) dx_k$ is a normalization factor. So far, we have established

the general analysis framework of probability density evolution filter (PDEF), which is illustrated in Fig.1.

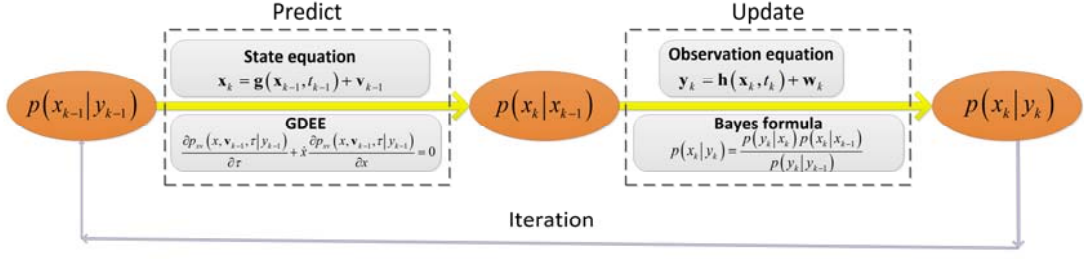

**Fig. 1 Schematic of probability density evolution filter**

Similar to PF, PDEF (Probability Density Evolution Filter) is capable of dealing with nonlinear dynamic systems with non-Gaussian noise. Whereas, there is generally no analytic solution to the GDEE (Generalized Density Evolution Equation), we have to sort to numerical methods to implement PDEF. Next, based on the pseudo-spectral method (collocation point method) for GDEE, PDEF can be sequentially implemented.

## 3. Numerical algorithm of probability density evolution filter (PDEF)

### 3.1 Pseudo-spectral scheme for generalized density evolution equation (GDEE)

In order to implement PDEF (Probability Density Evolution Filter), the key point is to solve the GDEE. Originally, a total variation diminishing (TVD) based finite difference scheme was developed by Li and Chen[10][14]. Theoretically, many numerical methods have been proposed to solve GDEE, e.g. [21][22][23][24]. In this work, we propose a collocation point based spectral method to obtain the numerical solution of GDEE[25].

Generally, the velocity term $\dot{x}$ in GDEE is time-varying. Whereas, it is reasonable to assume that $\dot{x}$ is a constant within a small enough time step $\Delta t = t_k - t_{k-1}$, written as $\dot{x}_k$. For simplicity, $p_{xv}(x, \mathbf{v}_{k-1}, \tau | y_{k-1})$ is abbreviated to $p$.

According to pseudo-spectral (collocation point) method, $p$ can be approximated with:

$$p(x,\tau) \approx \sum_{i=0}^{N} p_{i,k}(\tau) h_i(x) \tag{14}$$

where $p_{i,k}(\tau) = p(x_i, \tau)$ represent unknown node values of $p$ at $x_i$ and $\tau \in [t_k, t_{k+1}]$, and $h_i(x)$ are orthogonal basis functions [26].

Theoretically, we can adopt any complete orthogonal series. In this study, Chebyshev polynomials are employed to solve GDEE. Accordingly, $h_i(x)$ are defined as the Lagrange interpolating polynomials [26][27]:

$$h_i(x) = \frac{(-1)^{N+1+i}(1-x^2)T_N'(x)}{c_i N^2 (x - x_i)} \tag{15}$$

The first kind Chebyshev polynomials $T_j(x)$ defined in $[-1,1]$ can be explicitly expressed as [28][29]:

$$T_j(x) = \cos(j \arccos x), \quad j = 0, 1, \cdots, N \tag{16}$$

The Chebyshev polynomials satisfy with the orthogonal relation:

$$\int_{-1}^{1} T_i(x) T_j(x) \frac{1}{\sqrt{1-x^2}} = \frac{\pi}{2} \gamma_j \delta_{ij}, \quad \gamma_j = \begin{cases} 2 & j=0, \\ 1 & otherwise. \end{cases} \quad (17)$$

The derivative of Chebyshev polynomials could also be determined utilizing the recursion relationship:

$$T_j(x) = -\frac{1}{2(j-1)} T'_{j-1}(x) + \frac{1}{2(j+1)} T'_{j+1}(x) \quad (18)$$

In order to implement spectral methods, we need to determine the differentiation matrix. Taking the Gauss–Lobatto collocation points into account, the Chebyshev spectral differentiation matrix $\mathcal{D}_N$ are defined as follows[30][31]:

$$\begin{aligned}
\mathcal{D}_{00} &= -\frac{2N^2+1}{6}, \quad \mathcal{D}_{NN} = \frac{2N^2+1}{6}, \\
\mathcal{D}_{jj} &= \frac{-x_j}{2(1-x_j^2)}, \quad j = 1, \cdots, N-1, \\
\mathcal{D}_{ij} &= \frac{c_i}{c_j} \frac{(-1)^{i+j+N}}{(x_i - x_j)}, \quad i \neq j, c_i = \begin{cases} 2 & i = 0 \text{ or } N, \\ 1 & otherwise. \end{cases}
\end{aligned} \quad (19)$$

For more properties of Chebyshev polynomials, one can refer to[26][27] [28] [29].

Now, substitute the approximation expansion Eq.(14) into the GDEE Eq.(10). At collocation points $x_j$, we have

$$\sum_{i=0}^{N} \dot{p}_{i,k}(\tau) h_i(x_j) = -\dot{x}_k \sum_{i=0}^{N} p_{i,k}(\tau) h'_i(x_j), \quad j = 0,1,\cdots,N \quad (20)$$

Here, $\dot{p}_{i,k}$ is the derivative of $p_{i,k}$ with respect to $\tau$.

Considering $h_i(x_j) = \delta_{ij}$, GDEE at the collocation points can be expressed as:

$$\dot{p}_{j,k}(\tau) = -\dot{x}_k \sum_{i=0}^{N} p_{i,k}(\tau) h'_i(x_j), \quad j = 0,1,\cdots,N \quad (21)$$

Let $\mathbf{p} = (p_{0,k}(\tau), p_{1,k}(\tau), \cdots, p_{N,k}(\tau))^T$, the above GDEE at the collocation points can be written in a matrix form:

$$\dot{\mathbf{p}} = \mathcal{L}_k \mathbf{p} \quad (22)$$

where $\mathcal{L}_k = -\dot{x}_k \mathcal{D}_N(x_j)$, and $\mathcal{D}_N(x_j)$ is the Chebyshev differential matrix. Correspondingly, the initial condition $\mathbf{p}(t_{k-1}) = (p_{0,k-1}(t_{k-1}), p_{1,k-1}(t_{k-1}), \cdots, p_{N,k-1}(t_{k-1}))^T$, which is computed by interpolating $p|_{\tau=t_{k-1}} = \delta(x) p_v(\mathbf{v}_{k-1})$ at $x_j$,

According to the theory of character line, the matrix form GDEE has the sequential exponential solution[29]:

$$\mathbf{p}(t_k) = \mathbf{p}(t_{k-1}) e^{\Delta t \mathcal{L}_k} \quad (23)$$

In this way, we can predict the prior PDF $p(x_k|x_{k-1})$. The above matrix exponential equation is here calculated in terms of Padé approximation with scaling and squaring[32].

**3.2 Numerical algorithm of probability density evolution filter (PDEF)**

As mentioned before, in order to conduct PDEF we need to solve GDEE numerically. Herein, a pseudo-spectral based sequential procedure is presented as following.

**Algorithm**: Pseudo-spectral based probability density evolution filter

a) **Initialization**:

1) Generate mesh points $x_0^{(q)}$ and draw representative point set $\mathbf{v}_0^{(q)}, q = 1, 2, \cdots, N_{sel}$;

2) Interpolate the initial condition to obtain the starting vector $\mathbf{p}(t_0) = \delta(x) p_x(x_0^{(q)}) p_v(\mathbf{v}_0^{(q)})$.

b) **Prediction**:

**For k=1:n**

1) Substitute $\mathbf{v}_{k-1}^{(q)}$ into the state equation Eq.(2) to obtain $x_k^{(q)} = g(x_0^{(q)}, \mathbf{v}_{k-1}^{(q)})$ of interest;

2) Calculate the velocity term $\dot{x}_k^{(q)}$ based on the second-order central difference scheme according to $x_k^{(q)}$;

3) Scale $x_k^{(q)}$, $\dot{x}_k^{(q)}$ and $\mathbf{p}(t_{k-1})$ into [-1,1] in terms of the Chebyshev Gauss–Lobatto points $x_j = -\cos(\pi j / N)$ and calculate the Chebyshev spectral differentiation matrix $\mathcal{D}_N$;

4) Substitute the scaled $\dot{x}_k^{(q)}$ into GDEE Eq.(10), construct the matrix exponential operator $\mathcal{L}_k^{(q)} = -\dot{x}_k^{(q)} \mathcal{D}(x_j)$, and impose periodic boundary conditions, then obtain $p^{(q)}(x_j, t_k)$ by Chebyshev based pseudo- spectral method shown in Eq.(21).

5) Calculate the prior PDF $p(x_k|x_{k-1}) = \sum_{q=1}^{N_{sel}} p^{(q)}(x_j, t_k)$.

c) **Update**:

1) Calculate the posterior PDF according to Bayes formula $p(x_k|y_k) = \dfrac{q(y_k|x_k) p(x_k|x_{k-1})}{p(y_k|y_{k-1})}$;

2) Let $\mathbf{p}(t_{k-1}) = p(x_k|y_k) p_v(\mathbf{v}_k^{(q)})$, and $\mathbf{v}_{k-1}^{(q)} = \mathbf{v}_k^{(q)}, q = 1, 2, \cdots, N_{sel}$.

**End for**

**4. Numerical example**

In this section, in order to validate the PDEF algorithm, we herein consider a well-known illustrative example presented in[2][33][34][35][36]. Both the state equation and observation equation in this example exhibit severe nonlinearity. Meanwhile, plus and minus values of the state variable all satisfy the observation equation, which may induce a reverse estimation of the realistic statement. Moreover, the proposed method is compared to UKF and PF.

The state and observation equations are respectively written as[2]:

$$\mathbf{x}_k = \frac{\mathbf{x}_{k-1}}{2} + \frac{25\mathbf{x}_{k-1}}{1+\mathbf{x}_{k-1}^2} + 8\cos(1.2k) + \mathbf{v}_{k-1} \qquad (24)$$

$$\mathbf{y}_k = \frac{\mathbf{x}_k^2}{20} + \mathbf{n}_k \qquad (25)$$

where $\mathbf{v}_{k-1}$ and $\mathbf{n}_k$ are process and observation noise vectors, they are Gaussian random variables with zero mean and variances $Q_{k-1}=10$, $R_k=1$. For UKF and PF, one hundred particles with resampling at every time step were employed to conduct the filter algorithm. With the same number of sampling points and grid points, PDEF was implemented to assess the state. For comparison, a sample realization and the predicted values are presented in Fig.2. It can be seen that UKF, PF and PDEF are all effective to track the real state. While PF and PDEF show better performance than UKF.

In order to qualitatively evaluate the performance of PDEF, the root mean squared error (RMSE) is employed to measure the total error, which is defined as:

$$\text{RMSE} = \sqrt{\frac{1}{N}\sum_{k=1}^{N}(\hat{x}_k - x_k)^2} \qquad (26)$$

where $x_k$ denotes the true value at time $k$, $\hat{x}_k$ represents the predicted value at time $k$, $N$ is the total time steps. RMSEs for UKF, PF and PDEF averaged by 50 runs are listed in Table 1. Similarly, RMSE of EKF is indeed larger than ones of PF and PDEF, while the performance of PDEF is close to PF.

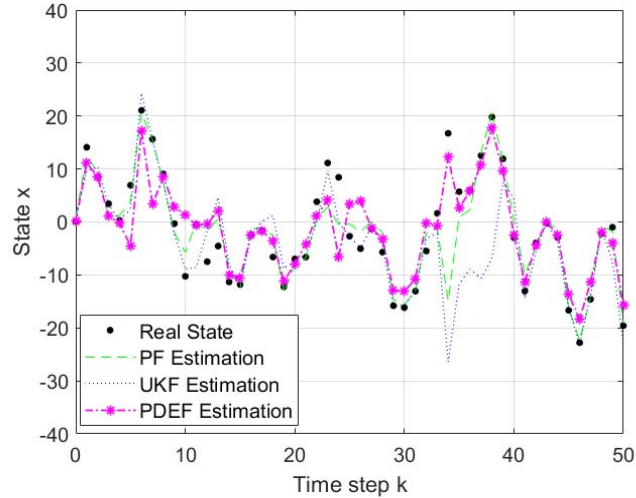

**Fig. 2** Comparison of a true state with the predicted values by UKF, PF and PDEF

**Table 1** Comparison of RMSE for different filter methods

| Algorithm | RMSE |
|---|---|
| Unscented Kalman Filter (UKF) | 7.04 |
| Particle Filter (PF) | 4.75 |
| Probability Density Evolution Filter (PDEF) | 5.62 |

## 5. Conclusions

A new filter framework is proposed based on probability density evolution theory and Bayes law. Focusing on only one component of all state variables and assuming that the rate of change of the system state is a constant within a small enough time step, the prior probability of the state component is predicted by means of the generalized density evolution equation (GDEE). Once a new observation information is available, the posterior probability of the state component is updated in terms of Bayesian formula. The proposed probability density evolution filter (PDEF) differs from particle filter (PF) in two ways. One is that PDEF evolve the prior probability in a differential form, in comparison to the integral Chapman–Kolmogorov equation in PF. Another is that there is no need of any resampling procedure in PDEF to predict the posterior probability. The numerical example shows good accuracy, close to PF and better than UKF.

**Acknowledgements:** This research is supported by the Natural Science Foundation of China (Grant No. 51578444), Key Science Research Program of Education Department of Shaanxi Province (20JY032).